\documentclass[conference]{IEEEtran}
\IEEEoverridecommandlockouts
\usepackage{cite}
\usepackage{amsmath,amssymb,amsfonts}
\usepackage{algorithmic}
\usepackage{graphicx}
\usepackage{textcomp}
\usepackage[dvipsnames]{xcolor}

\usepackage{hyperref}
\usepackage{orcidlink}

\usepackage{comment}
\usepackage{pgf-pie}
\usepackage{siunitx}
\usepackage[utf8]{inputenc}
\usepackage[T1]{fontenc}

\DeclareSIUnit{\dBm}{\text{dBm}}                            

\def\BibTeX{{\rm B\kern-.05em{\sc i\kern-.025em b}\kern-.08em T\kern-.1667em\lower.7ex\hbox{E}\kern-.125emX}}

\DeclareRobustCommand{\IEEEauthorrefmark}[1]{\smash{\textsuperscript{\footnotesize #1}}}

\usepackage{eso-pic}
\usepackage{url}
\AddToShipoutPictureBG*{
  \AtPageUpperLeft{%
    \put(0,-40){\raisebox{15pt}{\makebox[\paperwidth]{\begin{minipage}{21cm}\centering
      \textcolor{gray}{This article has been accepted for publication in the proceedings of the 2023 56th IEEE International Symposium on Circuits \\and Systems (ISCAS). Content may change prior to final publication. DOI: 10.1109/ISCAS46773.2023.10181431} 
    \end{minipage}}}}%
  }
  \AtPageLowerLeft{%
    \raisebox{25pt}{\makebox[\paperwidth]{\begin{minipage}{21cm}\centering
      \textcolor{gray}{ \copyright 2023 IEEE.  Personal use of this material is permitted.  Permission from IEEE must be obtained for all other uses, in any current or future media, including reprinting/republishing this material for advertising or promotional purposes, creating new collective works, for resale or redistribution to servers or lists, or reuse of any copyrighted component of this work in other works.
      }
    \end{minipage}}}%
  }
}

\begin{document}

\title{TinyBird-ML: An ultra-low Power Smart Sensor Node for Bird Vocalization Analysis and Syllable Classification \\
}


\author{\IEEEauthorblockN{ Lukas Schulthess\IEEEauthorrefmark{1},
Steven Marty\IEEEauthorrefmark{1},
Matilde Dirodi\IEEEauthorrefmark{2},
Mariana D. Rocha\IEEEauthorrefmark{2},}

\IEEEauthorblockN{
Linus Rüttimann\IEEEauthorrefmark{2},
Richard H. R. Hahnloser\IEEEauthorrefmark{2},
Michele Magno\IEEEauthorrefmark{1}}\\

\IEEEauthorblockA{\IEEEauthorrefmark{1}Dept. of Information Technology and Electrical Engineering, ETH Z\"{u}rich, Switzerland}
\IEEEauthorblockA{\IEEEauthorrefmark{2}Institute for Neuroinformatics,
ETH Z\"{u}rich, Switzerland}
}

\maketitle


\begin{abstract}
Animal vocalisations serve a wide range of vital functions. Although it is possible to record animal vocalisations with external microphones, more insights are gained from miniature sensors mounted directly on animals' backs. We  present TinyBird-ML; a wearable sensor node weighing only 1.4 g for acquiring, processing, and wirelessly transmitting acoustic signals to a host system using Bluetooth Low Energy. TinyBird-ML embeds low-latency tiny machine learning algorithms for song syllable classification.
To optimize battery lifetime of TinyBird-ML during fault-tolerant continuous recordings, we present an efficient firmware and hardware design. We make use of standard lossy compression schemes to reduce the amount of data sent over the Bluetooth antenna, which increases battery lifetime by 70\% without negative impact on offline sound analysis. Furthermore, by not transmitting signals during silent periods, we further increase battery lifetime. 

One advantage of our sensor is that it allows for closed-loop experiments in the microsecond range by processing sounds directly on the device instead of streaming them to a computer. 
We demonstrate this capability by detecting and classifying song syllables with minimal latency and a syllable error rate  of 7\%, using a light-weight neural network that runs directly on the sensor node itself. Thanks to our power-saving hardware and software design, during continuous operation at a sampling rate of 16 kHz, the sensor node achieves a lifetime of 25 hours on a single size 13 zinc-air battery.
\newline
\end{abstract}


\section{Introduction}\label{sec:Introduction}
Among all animal forms of communication, vocal expression contains much information about an individual's physiological and behavioral states~\cite{microphone_array, animal_acoustic, jurgens2001vocal, portfors2007types, seyfarth2003meaning}. Thus, to better understand animals' behaviours and their group dynamics, it is of utmost importance to study vocal patterns and their meanings~\cite{acoustic_activity_seal}. Unfortunately, the unambiguous assignment of a sound to the individual that generated it can be a major challenge, especially when animals rapidly move and irregularly vocalize in the midst of background sounds. A possible solution to meet this challenge is to record animals with wireless animal-borne sensors to directly capture vocalizations at their source~\cite{acoustic_monitoring_of_small_birds, control_vocal_interactions}.

Such animal-borne bio-loggers have been successfully used on large and mid-sized animals \cite{animal_acoustic, listening_to_lions, animal_iot_monitoring, on-animal_acoustic_recording}.
For small animals such as songbirds, in particular zebra finches (Taeniopygia guttata), the design of animal-borne sensor nodes is inflicted by many constraints \cite{brunecker2019tinybird, Pegan_2018}. 
Design imperatives are avoidance of stress from the weight or shape of the sensor node, because otherwise the behavioral readout may be of no value \cite{acoustic_monitoring_of_small_birds}. For this reason, bio-loggers need to be comfortable, very small and light-weight \cite{magno2019bluetooth}.
At the same time, to ensure an undisturbed observation period, a long battery lifetime is desired for chronic recordings. A day-long recording typically lasts 14 hours, which provides sufficient time for a battery change at night when birds do not sing and move~\cite{brunecker2019tinybird}.

Traditionally, animal-borne sensors are based on frequency modulation (FM) to transmit data \cite{reconstruction_of_vocal_interactions, zebra_finch_mates}. Although this approach can achieve minimal power consumption \cite{ruttimann}, there are well-known issues in terms of signal quality and interference when more than a single transmitter is in the same area \cite{indoor_radio}. Moreover, FM transmitters allow only one-way communication from the sensor to the receiver, and there is no way to send information back to the sensor node on the bird \cite{zebra_finch_mates}.

Recent technological advances in the fields of sensors, wireless communication, and power-efficient power MCUs Microcontrollers (MCUs) have paved the way for new approaches \cite{acoustic_monitoring_of_small_birds, brunecker2019tinybird, long-term_ble}. Switching from fully analog to digital solutions such as Bluetooth Low Energy (BLE) offers several advantages: communication is bidirectional and it allows the host station to send messages and commands to a sensor node, which is not possible with FM transmitters. Moreover, BLE allows simultaneous communication with multiple nodes and the on-board MCU is capable of performing complex mathematical operations such as running small neural networks to detect specific events in the data stream \cite{scherer2021tinyradarnn}. Such events in the field of bird vocalization analysis are the occurrence of syllables, which involves segmenting the syllables in time and identifying the syllable types~\cite{cohen2020tweetynet}. Moving from post-processing to real-time on-device syllable detection enables new research opportunities.

Since longitudinal observations are preferable and frequent battery changes can cause stress to an animal, animal-borne devices must be power-optimized on both hardware and firmware levels. Energy-efficient compression algorithms as described in \cite{multiterminal_source_coding} enable compression with tolerable data loss. On-board compression therefore can contribute to improving the energy efficiency by reducing the transmitted data volume~\cite{wireless_communication, wake-up_radio}. 

The paper presents TinyBird-ML, a miniaturized and power-optimized digital animal-borne sensor node that features a zero-power digital micro electromechanical system (MEMS) microphone and an inertial measurement unit (IMU). The node also hosts non-invasive electrodes in combination with a state-of-the-art electrostatic sensor to enable the acquisition of signals such as heart rate, respiration rate and muscle contractions which provides additional information for assessing the birds behaviour \cite{qvar_heartbeat, bird_vital_sign}.
The paper demonstrates the importance of on-board processing using tiny machine learning, proposing a light-weight neural network that fully runs on the TLE microcontroller and allows real-time signal analysis. We present a dual-stage classifier for detection of syllables and classification of their types. Due to the low-power design, including a "zero-power listening" acoustic sensor and on-board processing, TinyBird-ML achieves a lifetime of up to 25 h on a single 0.8 gram zinc-air battery. 

\section{System Architecture}\label{sec:System_Architecture}
\begin{figure}[htbp]
    \centering
    \includegraphics[width=\columnwidth]{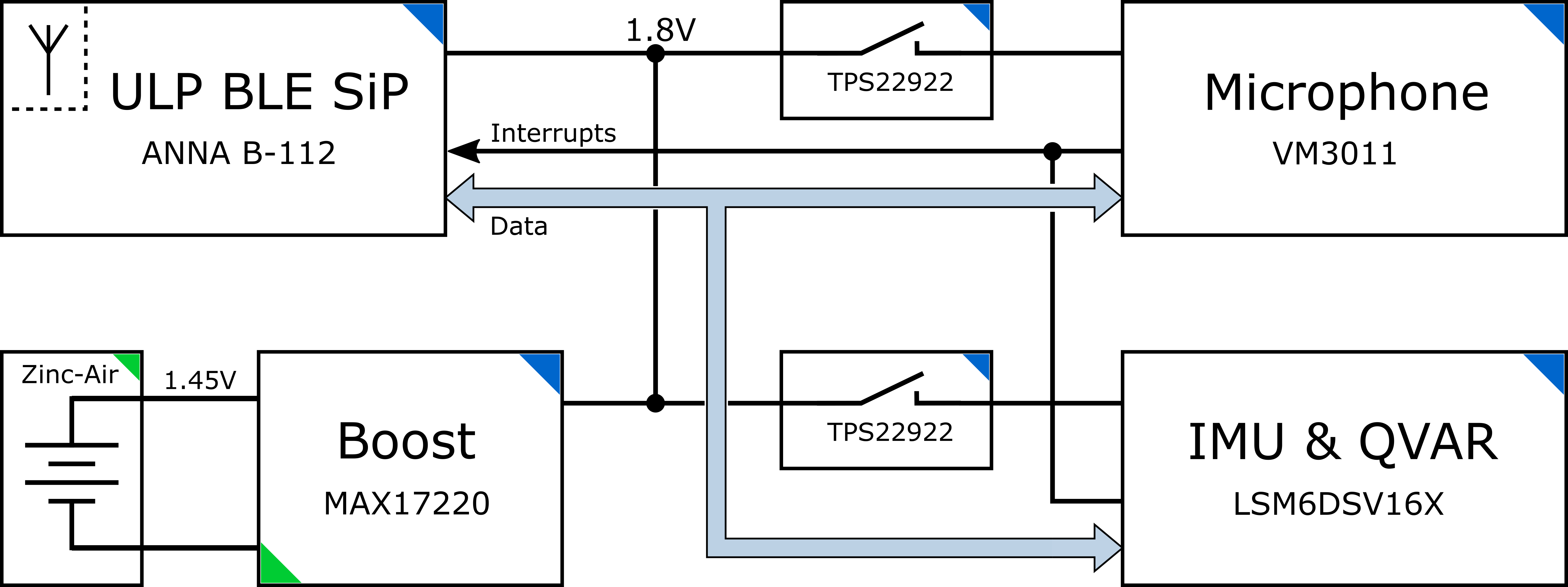}
    \caption{High-level block diagram of the sensor node.}
    \label{fig:tinybird_block_diagram}
\end{figure}

TinyBird-ML has been designed with low power and energy efficiency in mind to maximize operating time whilst keeping the total weight and therefore the impact on birds' natural behaviors at a minimum. Furthermore, the sensor nodes' elongated shape is designed to not disturb a bird during flight.
\autoref{fig:tinybird_block_diagram} shows a simplified block diagram of the proposed light-weight animal-borne sensor node for vocal monitoring. The system can be divided into three sub-parts:

\textit{\ref{sub:ble_module};} the core of the sensor node is the ANNA-B112 BLE, a module from ublox, \textit{\ref{sub:sensors};} the sensors for data acquisition and \textit{\ref{sub:power_management};} the power management based on a power-efficient boost-converter. Finally, the hardware design is described in \ref{sub:hardware_design}.

\subsection{Bluetooth Low Energy Module}\label{sub:ble_module}
The \textit{ANNA-B112} BLE module from ublox is an ultra compact System in Package (SiP) based on the nRF52832 System on Chip (SoC) from Nordic Semiconductor. It integrates Radio Frequency (RF) matching as well as an internal antenna. The ARM Cortex-M4F microcontroller with \SI{64}{\kilo\byte} RAM and \SI{512}{\kilo\byte} flash memory can run at a clock frequency up to \SI{64}{\mega\hertz} and thus offers enough computational power and memory to perform on-board audio processing and to run small neural networks. To increase the computational efficiency, the integrated Digital Signal Processor (DSP) can be exploited for compression algorithms. With a current consumption of \SI[per-mode = symbol]{58}{\micro\ampere\per\mega\hertz}
while running from flash memory, together with a Tx peak current of \SI{5.3}{\milli\ampere} at \SI{0}{\dBm}, the module offers outstanding performance in a small package. Its excellent trade-off between light weight, size, and low power consumption is the main reasons for selecting the ANNA-B112 BLE module.

\subsection{Sensors}\label{sub:sensors}
For sound acquisition, we used the high performance digital piezoelectric MEMS microphone \textit{VM3011} from Vesper MEMS.
It features adaptive zero-power listening, allowing it to reduce the microphone current consumption down to \SI{10}{\micro\ampere} while still being capable of detecting sounds louder than the background noise level. This functionality allows TinyBird-ML to further minimize the power consumption while the bird is not vocalizing, for example after the stressful sensor node attachment.
As soon as the bird has recovered and the desired behavior (e.g. song) is recognized, the microphone generates a wake-up signal and the microcontroller leaves its power-saving idle state. 

The \textit{LSM6DSV16X}, an IMU with integrated Qvar (Q = charge var = variation) electrostatic sensor from ST Microelectronics for non-invasive vital signs acquiring completes the sensor node. 
Both, MEMS microphone and IMU-Qvar sensor communicate over individual I2C peripherals and, depending on the application, can be power-gated using load switches.

\subsection{Power Management}\label{sub:power_management}
TinyBird-ML is powered by a non-rechargeable A13 standard-size \SI{1.45}{\volt} zinc-air battery, which is commonly used in hearing aids. Such batteries benefit from a very high theoretical energy density of \SI[per-mode = symbol]{1086}{\watt\hour\per\kilogram} 
\cite{zinc_air_battery}, which is crucial for our application. To overcome the low cell voltage, the nano-power synchronous boost converter \textit{MAX17220} from Maxim is
integrated. It converts the low input voltage to the sensor node's operating voltage of \SI{3}{\volt}. 
The boost converter's minimal input voltage of \SI{400}{\milli\volt} allows to completely deplete the battery and, together with a high conversion efficiency of over \SI{90}{\percent}, ensures an optimal power usage during operation.

\subsection{Hardware Design}\label{sub:hardware_design}
\begin{figure}[htbp]
    \centering
    \includegraphics[width=0.9\columnwidth]{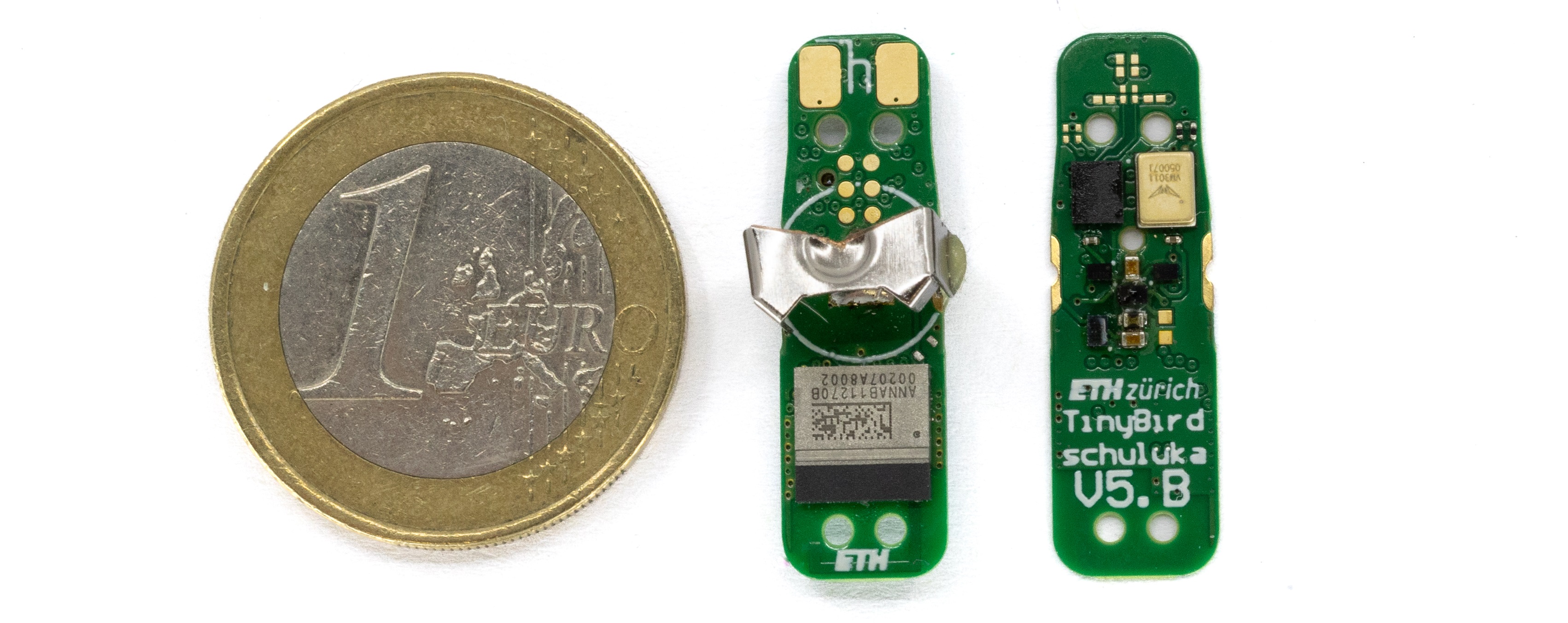}
    \caption{The space-, weight-, and power-optimized sensor node. The size of the body is only \SI{27.15}{\milli\meter} \(\times\) \SI{8.15}{\milli\meter}.
    Left: Top view of the sensor node with the electrode contacts on top and the BLE SoC at the bottom. Right: Bottom view of the sensor node with the Qvar top-left and the microphone top-right.}
    \label{fig:tinybird_hardware_top}
\end{figure}

\begin{figure}[htbp]
    \centering
    \begin{tikzpicture}[scale=1]
        \pie[
        text = legend,
        sum = auto,
        explode = 0.1,
        radius=1.6,
        color = {
            Goldenrod!50, 
            RoyalBlue!30, 
            ForestGreen!30,
            Gray!50,
            teal!30},
        ]
        {125/Microcontroller,
        101/Components \& solder,
        190/PCB,
        782/Battery,
        157/Battery clip}
    \end{tikzpicture}
    \caption{Weight distribution of the proposed sensor node in milligrams. Its total weight, including battery, amounts to \SI{1.4}{\gram}}
    \label{fig:weight_distribution}
\end{figure}
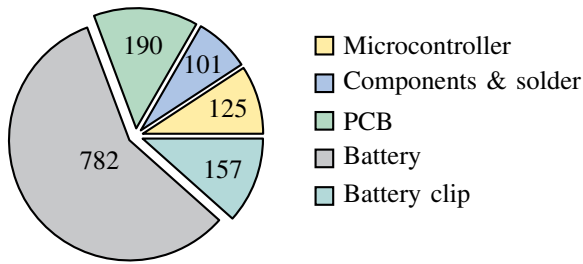

The entire sensor node is realized on a rigid 4-layer PCB with a total thickness of \SI{0.4}{\milli\meter}. The zinc-air battery is held by a weight-optimized battery clip, allowing for a reliable contact, facilitating battery replacement and achieving device reusability. Two contacts on the sensor node's top side can be used to attach two Qvar electrodes for cutaneous measurements of electrocardiograms (ref. \autoref{fig:tinybird_hardware_top}). 
For bird-borne sensor nodes, the overall weight is very crucial. The total weight of TinyBird-ML, including battery, amounts to \SI{1.4}{\gram}.
\autoref{fig:weight_distribution} summarizes the weight and size of each part.

\begin{figure}[htbp]
    \centering
    \includegraphics[width=0.4\columnwidth]{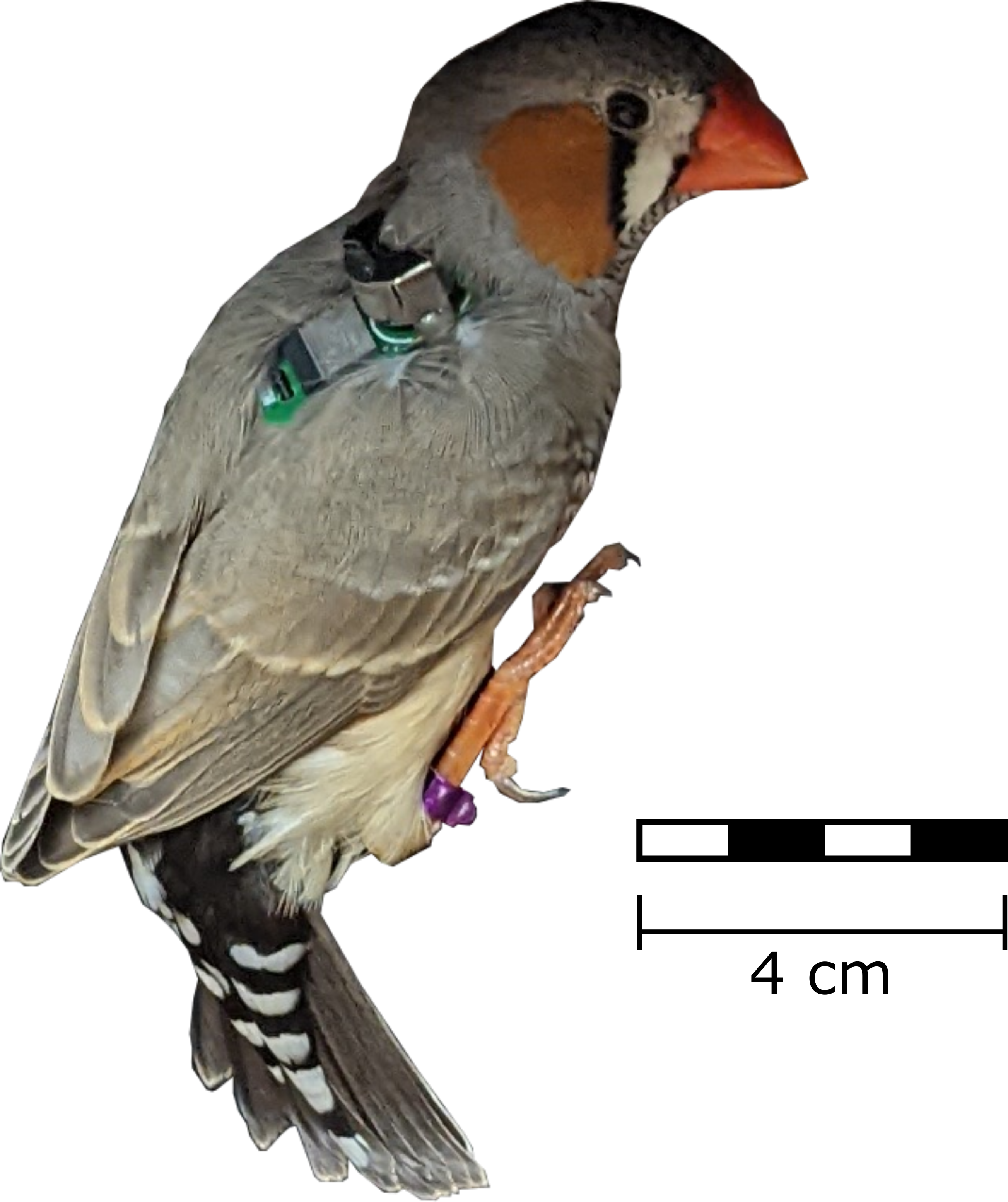}
    \caption{Male zebra finch with a mounted sensor node.}
    \label{fig:tinybird_mounted}
\end{figure}

\section{Firmware}\label{Firmware}

Once TinyBird-ML is powered up, it starts to send advertising packets. After having set up a connection, the central node subscribes to the TinyBird-ML's BLE characteristic responsible for the audio data. The recording does not start immediately because the bird may need time to acclimate itself to the backpack. To save power during this period, the BLE Connection Interval (CI) is set to a high value to save transmission power.

\subsection{Smart Compression}
In applications of vocal communication research, the main emphasis is on calls and song syllables. Therefore, to save energy, we tried to avoid transmission of sounds due to the movement of the bird and during silent periods altogether. Nevertheless, to allow for reconstruction of vocal output, we transmitted the temporal information about non-vocal periods. 
A simple approach is an amplitude threshold: acoustic signals below the threshold are classified as silent and lead to incrementation of a silence counter. Silent phases are thereby encoded as the counter magnitude. When the amplitude exceeds the threshold, the recording is started, compressed and send to the central, the silence counter is prepended to the audio block to allow correct reconstruction of the audio signal in the time domain.

\subsection{Syllable Classification using TinyML}
\begin{figure}[htbp]
    \centering
    \includegraphics[width=0.9\columnwidth]{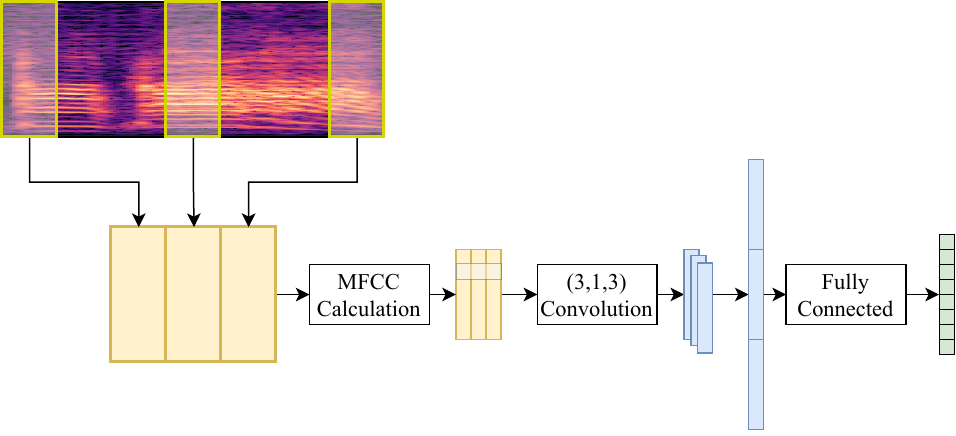}
    \caption{Syllable classification network.}
    \label{fig:syllable_classification}
\end{figure}

Zebra finch calls and song syllables can span over half a second, and the gaps between them can be as short as a few milliseconds \cite{araki2016mind, sasahara2015rhythm}. This indicates that no practical analysis window is small enough to always contain parts of just a single syllable and large enough to always include an entire syllable. To detect and analyze syllables, the data is windowed into non-overlapping blocks of a power-of-two samples, such that the Fast Fourier Transform (FFT) could be computed efficiently.
The signal blocks are processed by using two hierarchical classifiers. The first classifier analyses each block and decides, based on a single-layer perceptron with 16 Mel-frequency Cepstral Coefficients (MFCCs) as input, whether it contains parts of a syllable. This keeps the computational overhead minimal. 
The second classifier recognizes syllable types and is based on a mix of convolutional and fully connected layers, as illustrated in \autoref{fig:syllable_classification}. By waiting until the end of the syllable before classifying its type, more information about the syllable can be gathered. Three evenly distributed blocks are selected between syllable on- and offset. Each block is then transformed into sixteen MFCCs and used as input to the syllable classifier network. After nonlinear activation functions, the activations are flattened and fed through a fully connected layer with a softmax layer at the end.

\section{Experimental Results}\label{sec:Experimental_Results}

\subsection{Smart Compression}\label{sub:smart_compression}
\begin{figure}[htbp]
    \centering
    \includegraphics[width=\columnwidth]{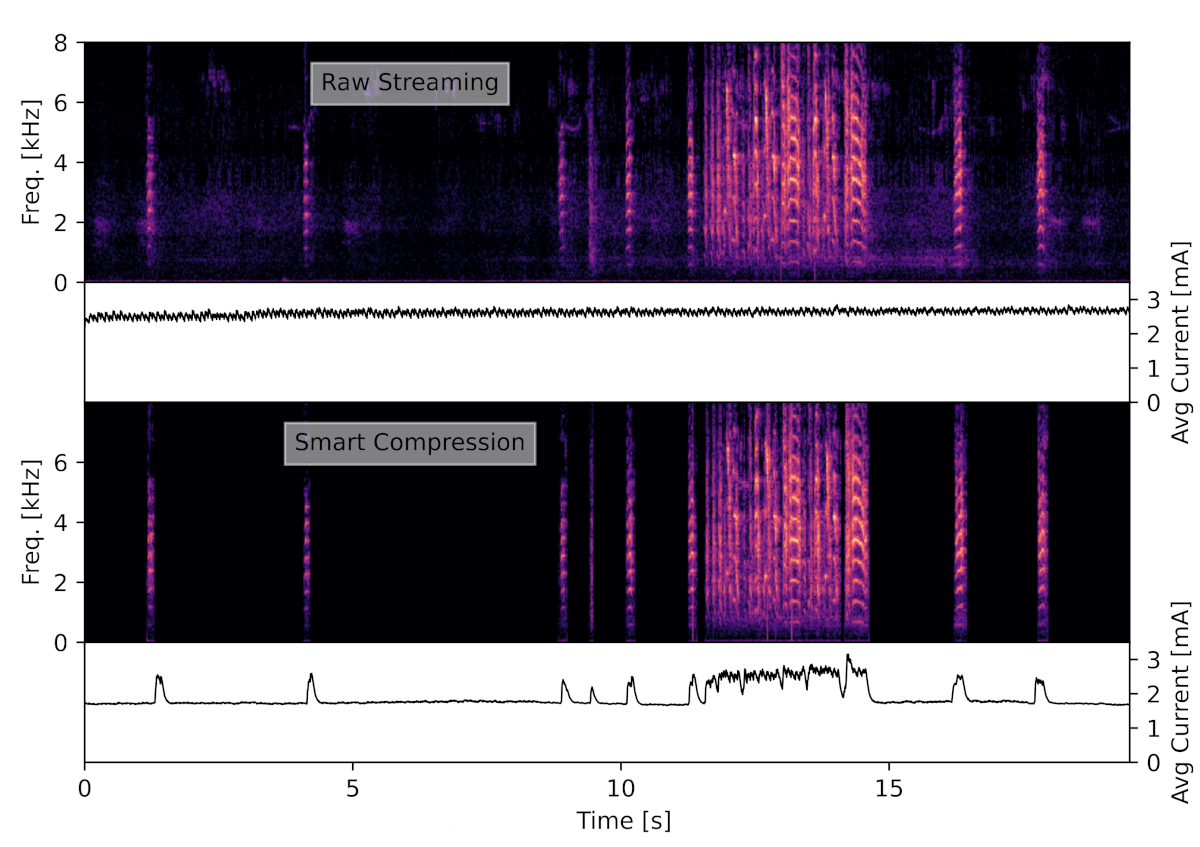}
    \caption{Spectrogram and average current consumption for raw data streaming and smart compression.}
    \label{fig:smart_compression}
\end{figure}

\autoref{fig:smart_compression} shows spectrograms of raw and of compressed audio signals. During reception of the latter, the current consumption only rises above the baseline level when vocalizations are detected and data is streamed over BLE. 
Comparing the two spectrograms reveals high-fidelity reconstruction from the compressed data.

\subsection{Compression Algorithms}\label{sub:compression_algorithms}
Compression algorithms reduce the number of bytes required to represent data at the cost of a computational overhead. To be effective, a compression scheme must achieve a reasonable quality wherein the energy savings from data reduction outweigh the power needed for compression. A comparison between different compression algorithms with respect to current consumption and memory usage is presented in \autoref{tab:algorithm_comparison}. Algorithms with a compression rate of 16 (Bitrate of \SI[per-mode = symbol]{2}{\kilo\bit\per\second}) represent the signals insufficiently for vocal analysis.
Among the remaining algorithms, the Adaptive Differential Pulse Code Modulation (ADPCM) shows the best performance with a minimal computational overhead of only \SI{50}{\micro\ampere} and a low average BLE current consumption of \SI{190}{\micro\ampere} 

\begin{table}[htbp]
   \centering
    \caption{Comparison of different compression algorithms.}
    \label{tab:algorithm_comparison}
    \resizebox{\columnwidth}{!}{
    \begin{tabular}{lcccccc} 
    \hline
    Compression & Bitrate & Overhead & Current & Overhead & Overhead   \\ 
    Algorithm & [\unit[per-mode = symbol]{\kilo\bit\per\second}] & [\unit{\milli\ampere}] & BLE [\unit{\milli\ampere}] & RAM [\unit{\kilo\byte}] & Flash [\unit{\kilo\byte}] \\
    \hline
    \hline
    Raw         & 32    & -      & 0.82     & -     & -     \\
    ADPCM       & 8     & 0.05   & 0.19     & < 1   & < 1   \\ 
    SBC High    & 8     & 0.11   & 0.20     & 1.5   & 5     \\ 
    Opus High   & 8     & 1.12   & 0.20     & 7     & 50    \\  
    DM          & 2     & -      & 0.06     & < 1   & < 1   \\ 
    CFDM        & 2     & 0.07   & 0.08     & <1    & < 1   \\ 
    SBC Low     & 2     & 0.07   & 0.07     & 1.5   & 5     \\ 
    Opus Low    & 2     & 0.87   & 0.09     & 7     & 50    \\
    \hline
    \end{tabular}
    }
\end{table}

\subsection{Syllable Classification}\label{sub:syllable_classification}
The proposed 8-bit quantized two-stage classification network achieved a syllable error rate~\cite{cohen2020tweetynet} of \SI{7}{\percent} on eight different syllable classes with an overall inference time of \SI{5.4}{\milli\second}.
\autoref{tab:classification_comparison} summarizes the classification networks performances and memory requirements.

\begin{table}[!t]
    \centering
    \caption{Performance of the classification network.}
    \label{tab:classification_comparison}
    \resizebox{\columnwidth}{!}{
    \begin{tabular}{lccc} 
    \hline
    Network & Inference & RAM Usage & Flash Usage   \\ 
    &  [\unit{\milli\second}] & [\unit{\kilo\byte}] & [\unit{\kilo\byte}] \\
    \hline
    \hline
    Detection       & 1.2   & 0.5   & 1.2    \\  
    Classification  & 4.2   & 1.2   & 2.7   \\
    \hline
    \end{tabular}
    }
\end{table}

\begin{figure}[htbp]
    \centering
    \includegraphics[width=\columnwidth]{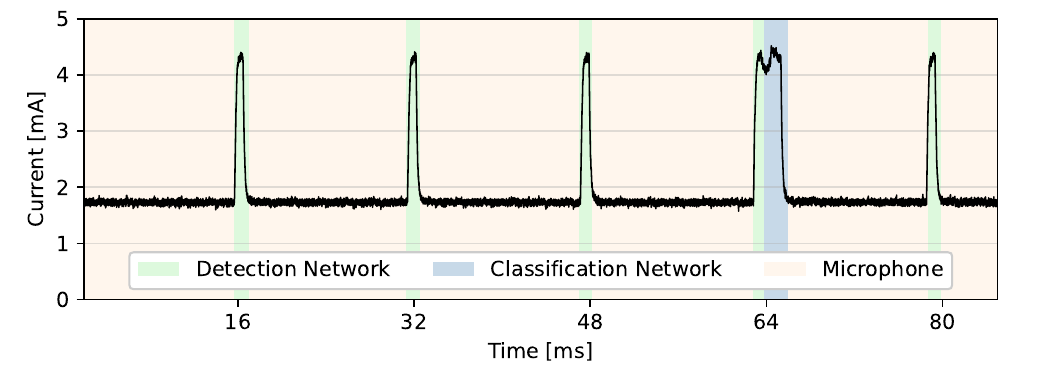}
    \caption{Current consumption for both detection and classification network, measured at \SI{3}{\volt}.}
    \label{fig:power_consumption_networks}
\end{figure}

For syllable detection, the microphone's buffer is fed through the detection network every \SI{16}{\milli\second}. The classification network only runs if a syllable has been detected (ref. \autoref{fig:power_consumption_networks}). Detected syllables sent over BLE require little data, including only the time between each syllable and the syllable type.
An average power consumption of \SI{5.73}{\milli\watt} has been achieved. It consists of the microphone's constant power consumption of \SI{5.25}{\milli\watt} and the classifier's computational overhead of \SI{0.48}{\milli\watt}.

The total power consumed is less than during audio streaming with any other standard compression algorithm, which makes on-board syllable detection an attractive power-saving solution.


\section{Conclusion}
This paper presented TinyBird-ML a miniaturized and power-optimized digital animal-borne sensor node that can be placed on small birds such as zebra finches. Besides a zero-power digital MEMS microphone, TinyBird-ML features an IMU with an integrated electrostatic sensor to enable monitoring a bird's vital signs such as its heart beats or muscle contractions. A light-weight neural network running on a microcontroller detects and classifies up to eight syllable types in real-time with a syllable error rate of 7\%. Weighing only 1.4 grams, TinyBird-ML minimizes weight whilst achieving an operating time of up to 25 hours during continuous data acquisition. 


\section*{Acknowledgment}
This work has been partially found by: the Swiss National Science Foundation (Agreement No 31003A-182638 and the NCCR Evolving Language Agreement no. 51NF40-180888), the BRAIN Initiative (Targeted BRAIN Circuits Project on ‘Neural sequences for planning and production of learned Vocalizations’), and the European Union’s Horizon 2020 Marie Sklodowska-Curie (grant 101025762 to M.D.R).

\newpage
\bibliographystyle{IEEEtran}
\bibliography{main}

\end{document}